\documentclass[%
 reprint,
superscriptaddress,
 amsmath,amssymb,
 aps,
]{revtex4-1}
\usepackage[version=3]{mhchem} 

\usepackage[left=1.5cm, right=1.5cm, top=1.785cm, bottom=2.0cm]{geometry}
\usepackage{balance}
\usepackage{times,mathptmx}
\usepackage{graphicx} 
\usepackage{lastpage}
\usepackage{amsmath,amssymb}
\usepackage{physics}
\usepackage{float}
\usepackage{fancyhdr}
\usepackage{fnpos}
\usepackage[english]{babel}
\usepackage{array}
\usepackage{droidsans}
\usepackage{charter}
\usepackage[T1]{fontenc}
\usepackage[usenames,dvipsnames]{xcolor}
\usepackage{soul}
\usepackage{setspace}
\usepackage[compact]{titlesec}
\usepackage{hyperref}
\usepackage{gensymb}
\usepackage{rotating}
\usepackage{multirow}
\usepackage{ulem}

\newcommand{\PP}{\mathrm{p}}
\newcommand{\NP}{\mathrm{np}}
\newcommand{\PS}{\mathrm{ps}}
\newcommand{\SD}{\mathrm{sd}}
\newcommand{\re}{$r_\mathrm{e}$}
\newcommand{\we}{$\omega_\mathrm{e}$}
\newcommand{\Ta}{$T_\mathrm{e}$}
\newcommand{\Tv}{$T_\mathrm{v}$}
\newcommand{\Lower}[1]{\smash{\lower 1.5ex \hbox{#1}}}

\setcitestyle{super}
\begin{document}
\title[Assessing the accuracy of simplified Coupled Cluster methods for electronic excited states in f0 actinide compounds]
 {Assessing the accuracy of simplified Coupled Cluster methods for electronic excited states in f0 actinide compounds}
\author{Artur Nowak}
\author{Pawe{\l} Tecmer}
\author{Katharina Boguslawski}
\email{k.boguslawski@fizyka.umk.pl}
\affiliation
{Institute of Physics, Faculty of Physics, Astronomy and Informatics, Nicolaus Copernicus University in Torun, Grudziadzka 5, 87-100 Torun, Poland}

 
 
\begin{abstract}
We scrutinize the performance of different variants of equation of motion coupled cluster (EOM-CC) methods to predict electronic excitation energies and excited state potential energy surfaces in closed-shell actinide species.
We focus our analysis on various recently presented pair coupled cluster doubles (pCCD) models [J.~Chem.~Phys., 23, 234105 (2016) and J.~Chem.~Theory~Comput. 15, 18--24 (2019)] and compare their performance to the conventional EOM-CCSD approach and to the completely renormalized EOM-CCSD with perturbative triples ansatz. 
Since the single-reference pCCD model allows us to efficiently describe static/nondynamic electron correlation, while dynamical electron correlation is accounted for \textit{a posteriori}, the investigated pCCD-based methods represent a good compromise between accuracy and computational cost.
Such a feature is particularly advantageous when modelling electronic structures of actinide-containing compounds with stretched bonds.
Our work demonstrates that EOM-pCCD-based methods reliably predict electronic spectra of small actinide building blocks containing thorium, uranium, and protactinium atoms.
Specifically, the standard errors in adiabatic and vertical excitation energies obtained by the conventional EOM-CCSD approach are reduced by a factor of 2 when employing the EOM-pCCD-LCCSD variant resulting in a mean error of 0.05 eV and a standard deviation of 0.25 eV.
\end{abstract}
\maketitle
\section{Introduction}
The efficient and reliable quantum chemical description of electronically excited states of atoms and molecules is of central importance in many areas of physics, chemistry, astrochemistry, and materials science.~\cite{bartlett_2007,Lischka1,kowalski-eom-2002,itoh-Fluorescence,Puzzarini1,gomes_rev_2012,Tennyson1,ola-pccp-2019,ola-book-chapter-2019} 
Despite the recent advances in computer architectures and the rapid progress in software optimization, there is still need for novel theoretical models that can overcome the exponential scaling wall of conventional CI-based methods. 
One promising alternative can be found in geminal-based approaches.
Specifically, geminal-based methods exploit an efficient parametrization of the electronic wavefunction using two-electron functions, also called geminals, which allows us to incorporate the dominant part of static/nondynamic electron correlation effects. 
Numerous numerical studies---ranging from simple model Hamiltonians, to non-covalent interactions, to bond-breaking processes including heavy elements---highlight that geminal-based ground-state electronic wavefunctions can reliable account for static/nondynamic electron correlation effects at low computational cost.~\cite{pernal2012,pawel_jpca_2014,oo-ap1rog,ps2-ap1rog,pernal2014,piotrus_mol-phys,ap1rog-jctc,pawel_pccp2015,boguslawski2016,pernal-2018,filip-jctc-2019}
This feature makes them ideal candidates to efficiently and reliably target excited states.
Particularly interesting are the Antisymmetric Product of Strongly orthogonal Geminals (APSG)~\cite{surjan_1999,surjan2012} and the pair Coupled Cluster Doubles (pCCD)~\cite{limacher_2013,oo-ap1rog,tamar-pcc} models (also known as the Antisymmetric Product of 1-reference orbital Geminal~\cite{limacher_2013,oo-ap1rog}) and their extensions to excited states.~\cite{pernal2012,pernal2014,pernal_jcp_2014,pernal2014_jcp_erratum,eom-pccd,eom-pccd-erratum,eom-pccd-lccsd}

Specifically, in the APSG ansatz, excited states have been modelled using different flavours of Extended Random Phase Approximation (ERPA)~\cite{pernal2012} and the time-dependent linear response formalism.~\cite{pernal_jcp_2014,pernal2014_jcp_erratum}
These approaches have been tested for excitation energies in Be, \ce{H_2}, LiH, \ce{N_2}, \ce{Li_2}, BH, \ce{H_2O}, and \ce{CH_2O}.~\cite{pernal2012,pernal2014,pernal_jcp_2014,pernal2014_jcp_erratum} 
In the pCCD model, excited states are targeted using the Equation-of-Motion (EOM) formalism.~\cite{eom-pccd,eom-pccd-erratum,eom-pccd-lccsd}
Two different excited state models have been presented in the literature that employ a pCCD reference functions and that can describe at most double excitations.
In the simplest variant, the EOM formalism is directly applied on top of pCCD. 
The missing single excitations are then included \textit{a posteriori} in the EOM ansatz, that is, the linear excitation operator of the EOM formalism is limited to pair and single excitations.
This approach is labeled as EOM-pCCD+S.~\cite{eom-pccd,eom-pccd-erratum}
A more sophisticated scheme uses the pCCD model with an \textit{a posteriori} linearized Coupled Cluster Singles and Doubles correction~\cite{kasia-lcc} (pCCD-LCCSD) as the reference function in the EOM model, resulting in the EOM-pCCD-LCCSD approach.~\cite{eom-pccd-lccsd}

The performance of the recently presented pCCD-based excited state models has been assessed for various challenging systems including \ce{CH^+}, all-trans polyenes, and heavy-element-containing compounds that feature several (quasi-)degenerate one-electron states like the ytterbium dimer~\cite{pawel-yb2} and the uranyl cation (\ce{UO_2^{2+}}).~\cite{eom-pccd,eom-pccd-erratum} 
Recent numerical studies suggest that EOM-pCCD-based methods predict excitation energies for singly-excited states that are comparable to EOM-CCSD results, while the excitation energies of doubly-excited states, that is, electronic excitation energies associated with a simultaneous transfer of two electrons, are closer to results obtained from multi-reference calculations.
However, a detailed analysis of the performance of EOM-pCCD-based methods for different types of excitation energies, like local, charge-transfer, or Rydberg excitations that can be encountered in various chemical systems across the periodic table, is missing.     
Furthermore, it is particularly important to benchmark alternative quantum chemistry methods for challenging molecules that feature complex electronic structures and electronic spectra like transition-metal, lanthanide, and actinide-containing compounds. 
This is especially desirable for larger heavy-element-containing species for which standard quantum chemistry methods are technically limited or computationally infeasible, while experimental manipulations are hampered due to the acute toxicity or instability of such compounds. 

To that end, the goal of this work is to assess the accuracy of various EOM-pCCD-based excited state extensions for a test set of f0 actinide species that represent small building blocks of larger actinide-containing compounds.
Their performance is compared to conventional Equation-of-Motion Coupled Cluster Singles Doubles~\cite{bartlett-eom} (EOM-CCSD) and to the Completely Reonormalized Equation of Motion Coupled Cluster Singles Doubles and perturbative Triples (CR-EOM-CCSD(T)) method.~\cite{cr-eomccsd} 
{We should stress that the CR-EOM-CCSD(T) approach has been successfully applied to model excited states of actinide species and can be considered as a reference method for closed-shell actinide containing species.
~\mbox{\cite{CR-EOMCCSDT_actinide_1,CR-EOMCCSDT_actinide_2,pawel_saldien,CR-EOMCCSDT_actinide_3}}
}
Specifically, we focus on the ThO and ThS diatomics and on the uranyl cation and its isoelectronic series containing the \ce{NUO^+}, \ce{NUN}, CUO, \ce{ThO_2}, and \ce{PaO_2^+} compounds. 
Such small, prototypical di- and tri-atomic f0 actinide-containing species represent helpful and robust models for predicting bonding mechanisms, electronic structures, and properties of larger realistic actinide compounds for which theoretical investigations are rather limited.~\cite{denning_91b,straka2001,denning2007}

This work is organized as follows. In section~\ref{sec:theory}, we briefly recapitulate all recently presented pCCD-based excited state methods. 
Section~\ref{sec:comp} summarizes the computational details. 
All predicted electronic spectra and their spectroscopic constants and statistical analysis are presented in section~\ref{sec:results}.
Finally, we conclude in section~\ref{sec:conclusions}.
\section{pCCD-based models for ground and excited states}\label{sec:theory}
Restricting the coupled cluster ansatz~\cite{coester_1958,cizek_jcp_1966,cizek_paldus_1971,paldus_cizek_shavitt_1972,bartlett_rev_1981,bartlett_2007} to include only pair-excitations in the cluster operator results in the pCCD approach,~\cite{limacher_2013,oo-ap1rog,tamar-pcc} 
\begin{equation}\label{eq:pccd}
|\textrm{pCCD} \rangle = \exp \big(\sum_{i=1}^P \sum_{a=P+1}^K t_i^a a^\dagger_a a^\dagger_{\bar{a}} a_{\bar{i}} a_i \big) | \Phi \rangle
                       = \exp (\hat{T}_{\mathrm{p}}) \ket{\Phi},
\end{equation}
where $|\Phi \rangle$ is some independent-particle wavefunction (for instance, the Hartree--Fock (HF) determinant), $a_p^\dagger$ ($ a_{\bar{p}}^\dagger$) and $a_p$ ($a_{\bar{p}}$) are the electron creation and annihilation operators, respectively, for electrons with $\alpha$ ($\beta$) spin. 
Numerical studies indicate that the pCCD model can be considered as an efficient way of capturing static/nondynamic electron correlation effects, provided the orbitals are optimized.~\cite{limacher_2013,oo-ap1rog,pawel_jpca_2014,tamar-pcc,piotrus_mol-phys,frozen-pccd,ps2-ap1rog,ap1rog-jctc,pawel_pccp2015,boguslawski2016}
Furthermore, a variational orbital optimization scheme outperforms various approximate orbital optimization protocols based on the seniority number.~\cite{ps2-ap1rog,ap1rog-jctc}

The missing fraction of the dynamic electron correlation energy can be accounted for~\textit{a posteriori} using, for instance, perturbation theory~\cite{piotrus_pt2,AP1roG-PTX} or (approximate) CC corrections on top of pCCD.~\cite{frozen-pccd,kasia-lcc} 
In the latter approach, an exponential ansatz for the electronic wavefunction is exploited,
\begin{equation}\label{eq:cc-ansatz}
|\Psi \rangle=  \begin{rm}{exp}\end{rm}(\hat{T}) |\textrm{pCCD} \rangle,
\end{equation} 
where $\hat{T}=\sum_{\nu}t_{\nu}\hat{\tau}_{\nu}$ is a general cluster operator that contains different levels of excitations $\hat{\tau}$. 
If single and double excitations with respect to the reference determinant are included in eq.~\eqref{eq:cc-ansatz}, we arrive at the frozen-pair CCSD (fpCCSD) model.~\cite{frozen-pccd}
The fpCCSD ansatz can be further simplified by considering only linear terms in the Baker-Campbell-Hausdorff expansion, which results in the pCCD-LCCSD method.~\cite{kasia-lcc}
The corresponding amplitudes are then determined by solving a set of linear equations,
\begin{equation}\label{eq:lcc-schrodinger}
	\langle\Phi_\nu|(\hat{H}_N+[\hat{H}_N, \hat{T}_{\mathrm{np}}])|\textrm{pCCD} \rangle = 0,
\end{equation}
where $\hat{H}_N=\hat{H}-\langle\Phi|\hat{H}|\Phi\rangle$ is the quantum chemical Hamiltonian in its normal product form, $\bra{\Phi_\nu}$ is a Slater determinant contained in the projection manifold, and $\hat{T}_{\mathrm{np}}$ contains all possible electron excitations up to second order excluding all electron-pair excitations.
Eq.~\eqref{eq:lcc-schrodinger} can be rewritten into the familiar form of single-reference coupled cluster theory,~\cite{kasia-lcc}
\begin{equation}\label{eq:lcc-pccd}
	\bra{\Phi_\nu}(\hat{H}_N+[\hat{H}_N, \hat{T}_{\mathrm{np}}])+[\hat{H}_N, \hat{T}_{\mathrm{p}}]+[[\hat{H}_N, \hat{T}_{\mathrm{np}}],\hat{T}_{\mathrm{p}}]) \ket{\Phi} = 0.
\end{equation} 
Various numerical studies highlight the good performance of the pCCD-LCCSD approach for many molecular systems, including heavy-element chemistry and non-covalent interactions, where both static and dynamic electron correlation effects are important.~\cite{kasia-lcc,AP1roG-PTX,pawel-yb2,filip-jctc-2019}  

\subsection{Targeting excited states within pCCD}\label{sec:ext}
For pCCD-based methods, excited states have been modelled using the EOM formalism.~\cite{eomcc_1968,eomcc_1989,kowalski-eom-review-2011,krylov-review}
In the standard EOM approach, the excited state wavefunction $\ket{\Psi_k}$ is approximated by a CI-type ansatz, $\hat{R}_k = \sum_{k} c_k \hat{\tau}_k$, acting on the corresponding coupled cluster reference state $\ket{\Psi}$,
\begin{equation}\label{eq:EOM-ansatz}
	|\Psi_k \rangle=  \hat{R}_k |\Psi \rangle= \sum_{k} c_k \hat{\tau}_k|\Psi \rangle. 
\end{equation} 
In the above equation, the sum runs over all excitations included in the cluster operator $\hat{T}$ of the coupled cluster reference wavefunction. 
Eq.~\eqref{eq:EOM-ansatz} is solved by determining the amplitudes of the $\hat{R}$ operator. 
The excited state in question is obtained by solving the corresponding Schr\"odinger equation,
\begin{equation}\label{eq:EOM-CC-schrodinger}
	\hat{H}_N|\Psi_k\rangle = \hat{H}_N\hat{R}_k|\Psi\rangle =   E_k\hat{R}_k|\Psi\rangle = E_k|\Psi_k\rangle, 
\end{equation}
where $E_k$ is the total electronic energy associated with excited state $k$.
Introducing the similarity transformed Hamiltonian $\hat{\mathcal{H}}_N=\exp(-\hat{T})\hat{H}_N\exp(\hat{T})$ and subtracting the Schr\"odinger equation of the coupled cluster ground state, we obtain the well-known EOM equations,
\begin{equation}\label{eq:EOM-CC}
	[\hat{\mathcal{H}}_N,\hat{R}_k]|\Phi \rangle=\omega_k \hat{R}_k|\Phi\rangle, 
\end{equation}
where $\omega_k =(E_k-E_0)$ are the excitation energies with respect to the CC ground state.
Solving eq.~\eqref{eq:EOM-CC} is equivalent to determine the eigenvalues and eigenvectors of a non-Hermitian matrix.

\subsection{EOM-pCCD}
The simplest excited state extension for pCCD is the EOM-pCCD model, where the cluster operator and hence the CI-ansatz of eq.~\eqref{eq:EOM-ansatz} is restricted to electron-pair excitations only.  
Specifically, we have
\begin{align}\label{eq:Rp-operator}
\hat{R}_\PP &= \hat{R}_0 + \hat{R}_\PP \nonumber \nonumber \\
            &= c_0\hat{\tau}_0 + \sum_{ia}c_{i\bar{i}}^{a\bar{a}}\hat{\tau}_{a\bar{a}i\bar{i}}. 
\end{align}
The corresponding EOM-pCCD equations, 
\begin{equation}\label{eq:EOM-pCCD}
[\hat{\mathcal{H}}_N^{(\PP)},\hat{R}_\PP]|\Phi \rangle=\omega_\PP \hat{R}_\PP|\Phi\rangle, 
\end{equation}
can thus only describe electron-pair excitations.
In the above equation, $\hat{\mathcal{H}}_N^{(\PP)}$ is the similarity transformed Hamiltonian of pCCD,  
\begin{equation}\label{eq:pair-Hamiltonian}
\hat{\mathcal{H}}_N^{(\PP)}=e^{-\hat{T}_\PP}\hat{H}_Ne^{\hat{T}_\PP} = \hat{H}_N  + [\hat{H}_N, \hat{T}_{\PP}] +{1\over{2}}[[\hat{H}_N, \hat{T}_{\PP}],\hat{T}_{\PP}] + \ldots
\end{equation}
To target singly-excited or general doubly-excited states, the EOM-pCCD model has to be extended.
Different flavours of post-EOM-pCCD methods have been presented in the literature that can be understood as various approximations of the conventional EOM-CCSD approach.~\cite{eom-pccd,eom-pccd-erratum,eom-pccd-lccsd}

\subsection{Extending EOM-pCCD to target singly-excited states: EOM-pCCD+S and EOM-pCCD-CCS}
In the simplest extension of EOM-pCCD, single excitations are introduced analogously to the configuration interaction method restricted to single excitations (CIS).
In this approach, only the $\hat{R}$ operator of eq.~\eqref{eq:Rp-operator} is modified and also includes single excitations with respect to the pCCD reference determinant,
\begin{align}\label{eq:Rps-operator}
\hat{R}_\PS&= \hat{R}_0 + \hat{R}_1 + \hat{R}_\PP \nonumber \\
           &= c_0\hat{\tau}_0 +\sum_{ia}c_{i}^{a}\hat{\tau}_{ai} +\sum_{ia}c_{i\bar{i}}^{a\bar{a}}\hat{\tau}_{a\bar{a}i\bar{i}},
\end{align}
where $\hat{\tau}_{ai}$ is a single excitation operator generating all singly excited states with respect to the pCCD reference determinant, $|^a_{i}\rangle=\hat{\tau}_{ai}|\Phi\rangle$.
The corresponding excitation energies and excited state wavefunctions are obtained by solving the corresponding EOM equations,
\begin{equation}\label{eq:EOM-pCCD+S}
[\hat{\mathcal{H}}_N^{(\PP)},\hat{R}_\PS]|\Phi \rangle=\omega_\PS \hat{R}_\PS|\Phi\rangle.  
\end{equation}
We should emphasize that in the above equation $\hat{\mathcal{H}}_N^{(\PP)}$ is the similarity transformed Hamiltonian of pCCD as defined in eq.~\eqref{eq:pair-Hamiltonian}.
Since single excitations have been introduced \textit{a posteriori}, the corresponding excited state model looses size-intensivity~\cite{eom-pccd,eom-pccd-erratum} and the first column of the EOM effective Hamiltonian does not equal zero,
\begin{equation}\label{eq:Matrix_1}
\left[ \begin{array}{ccc}
0       &      \langle 0|\hat{\mathcal{H}}_N^{(\PP)}|^b_{j}\rangle         & \langle 0|\hat{\mathcal{H}}_N^{(\PP)}|^{b\bar{b}}_{j\bar{j}}\rangle \\
\langle ^a_{i}|\hat{\mathcal{H}}_N^{(\PP)}|0\rangle   & \langle ^a_{i}|\hat{\mathcal{H}}_N^{(\PP)}|^b_{j}\rangle   & \langle ^a_{i}|\hat{\mathcal{H}}_N^{(\PP)}|^{b\bar{b}}_{j\bar{j}}\rangle  \\
0       & \langle ^{a\bar{a}}_{i\bar{i}}|\hat{\mathcal{H}}_N^{(\PP)}|^b_{j}\rangle & \langle ^{a\bar{a}}_{i\bar{i}}|\hat{\mathcal{H}}_N^{(\PP)}|^{b\bar{b}}_{j\bar{j}}\rangle
\end{array} \right]. 
\end{equation}
The size-intensivity error can be straightforwardly cured by modifying the cluster operator of the coupled cluster reference wavefunction.
If we keep the EOM model restricted to single and electron-pair excitations, we have to perform an \textit{a posteriori} CCS calculation on top of pCCD (\textit{cf.}, eq.~\eqref{eq:cc-ansatz} with $\hat{T}=\hat{T}_1$, abbreviated as pCCD-CCS).
The corresponding EOM equations,
\begin{equation}\label{eq:EOM-pCCD-CCS}
[\hat{\mathcal{H}}_N^{(\PS)},\hat{R}_\PS]|\Phi \rangle=\omega_\PS \hat{R}_\PS|\Phi\rangle.  
\end{equation}
contain the similarity transformed Hamiltonian of pCCD-CCS $\hat{\mathcal{H}}_N^{(\PS)}$.
Since single excitations are now accounted for in the coupled cluster reference wavefunction, the first column in eq.~\eqref{eq:Matrix_1} equals zero.
We should note that the similarity transformed Hamiltonian used in the EOM-pCCD-CCS equations has a different form for electron-pair and single excitations, respectively.
While for all electron-pair excitations $\hat{\mathcal{H}}_N^{(\PS)}$ has the same expression as $\hat{\mathcal{H}}_N^{(\PP)}$ defined in eq.~\eqref{eq:pair-Hamiltonian},
for single excitations the cluster operator contains both single and electron pair excitations,
\begin{equation}\label{eq:nonpair-Hamiltonian}
\hat{\mathcal{H}}_N^{(\PS)}=e^{-\hat{T}_1 - \hat{T}_\PP}\hat{H}_Ne^{\hat{T}_1 + \hat{T}_\PP}.
\end{equation}

\subsection{Targeting singly and general doubly-excited states: EOM-pCCD-LCCSD}
By construction, general doubly-excited states are not accessible in EOM-pCCD+S nor EOM-pCCD-CCS.
To target general bi-excited states, post-pCCD methods have to be considered that contain---at least---the $\hat{T}_2$ excitation operator in the CC correction.
So far, only the pCCD-LCCSD method has been extended to model excited states.~\cite{eom-pccd-lccsd}
Since the cluster operator of pCCD-LCCSD contains both single and double (that is, both electron-pair and broken-pair) excitations, the $\hat{R}$ operator has the general form,
\begin{align}\label{eq:Rsd-operator}
\hat{R}_\SD&=\hat{R}_0+\hat{R}_1+\hat{R}_2 \nonumber \\
           &= c_0\hat{\tau}_0 +\sum_{ia}c_{i}^{a}\hat{\tau}_{ai} + \frac{1}{2}\sum_{ijab}c_{ij}^{ab}\hat{\tau}_{abij}.
\end{align}
Substituting the above expression for the $\hat{R}$ operator in the general EOM equations~\eqref{eq:EOM-CC}, the EOM-pCCD-LCCSD equations become 
\begin{equation}\label{eq:EOM-pCCD-LCCSD}
[\hat{\mathcal{H}}_N^{(\SD)},\hat{R}_\SD]|\Phi \rangle=\omega_\SD \hat{R}_\SD|\Phi\rangle.
\end{equation}
Similar to EOM-pCCD-CCS, different expressions for the similarity transformed Hamiltonian associated with electron-pair and non-electron-pair excitations are used.
Specifically, for non-electron-pair excitations, we have~\cite{eom-pccd-lccsd}
\begin{equation}\label{eq:nonpair-Hamiltonian}
	\hat{\mathcal{H}}_N^{(\NP)}=e^{-\hat{T}_\PP - \hat{T}_{\NP}}\hat{H}_Ne^{\hat{T}_\PP + \hat{T}_{\NP}}\approx \hat{H}_N  + [\hat{H}_N, \hat{T}_{\NP}] +[[\hat{H}_N, \hat{T}_{\NP}],\hat{T}_{\PP}].
\end{equation}

To conclude, the exponential ansatz of pCCD-based methods ensures size-extensivity of the model.
Size-consistency errors may, however, become problematic, especially when moving along potential energy surfaces (PES) (for stretched bonds or in the vicinity of dissociation).
The quality of the PES further determines the accuracy of spectroscopic constants.
To remedy size-consistency problems, the orbital basis has to be optimized, ideally for each targeted excited state.
In this work, we have performed orbital optimizations for the pCCD ground state wavefunction only.
Although such a procedure cures size-consistency errors, the optimized orbitals might be symmetry-broken, which complicates the identification of excited states.
Symmetry-breaking can be prevented by imposing point group symmetry during the orbital optimization (see also sec.~\ref{sec:comp} for more details). 

\subsection{Computational scaling}
{Finally, we would like to compare the computational scaling of different EOM-CC-based methods considered in this work. 
The computational cost of EOM-CCSD is comparable to CCSD (requires, however, additional disk space) and scales as $\mathcal{O}(o^2v^4)$, where $o$ is the number of occupied orbitals and $v$ is the number of virtual orbitals.
The Scalability of the triples correction of the CR-EOM-CCSD(T) method is characterized by $\mathcal{O}(n^7)$ numerical complexity, where $n$ is the number of basis functions.\mbox{\cite{kowalski2010active}}
The simplest and simultaneously cost-effective EOM-pCCD-based models scale as $\mathcal{O}(o^2v^2)$. 
The inclusion of single excitations within the EOM-pCCD+S and EOM-pCCD-CCS framework does not significantly increase the computational cost compared to EOM-pCCD and contains only a larger pre-factor. 
The computational cost of EOM-pCCD-LCCSD is comparable to the standard EOM-CCSD formalism.\mbox{\cite{eom-pccd,eom-pccd-lccsd}}
Finally, we should mention that the EOM-pCCD-based methods can be combined with an orbital optimization protocol within pCCD, which introduces an additional cost of several 4-index transformations (formally $\mathcal{O}(n^5)$).
A short summary of CPU timings for different excited state models is presented in Table S22 of the ESI\dag}
\section{Computational details}\label{sec:comp}

\subsection{Basis sets, relativity, and frozen core}
For ThO and ThS, we have used the all-electron atomic natural orbital relativistic correlation consistent (ANO-RCC) basis sets~\cite{ano-rcc_u} available in the OpenMolcas program package version 17.0,~\cite{molcas8} optimized specifically for the 2-nd order Douglas--Kroll--Hess (DKH2) Hamiltonian.~\cite{dkh1, dkh2,reiher_book,tecmer2016}
For all remaining molecules, we have employed the double-$\zeta$ correlation consistent basis sets of Peterson~\cite{cc-pvdz-dk3-actinides} for all heavy elements (cc-pVDZ-DK3), optimized specifically for the DKH3 Hamiltonian,~\cite{reiher_2004a,reiher_2004b,reiher_book,tecmer2016} and Dunning's aug-cc-pVDZ basis set for all light elements.~\cite{basis_dunning} 
Scalar relativistic effects were accounted for by the DKH2 Hamiltonian (ThO and ThS) and the DKH3 Hamiltonian (\ce{UO2^2+} and its isoelectronic series), respectively.
We should note that we performed additional test calculations with basis sets of triple-$\zeta$ quality.
Since all investigated EOM-CC methods yield similar excitation energies for a double-$\zeta$ and triple-$\zeta$ quality basis set (differences are typically 0.1 eV or smaller), all calculations have been performed with a double-$\zeta$ quality basis set.
Results obtained for the corresponding triple-$\zeta$ quality basis set are summarized in section S8 of the ESI\dag.

{In all coupled cluster calculations, the atomic 1s orbitals of the C, O, and N atom, the 1s--2p orbitals of the S atom, and the 1s--5d orbitals of the Th, Pa, and U center were kept frozen. 
This choice represents a compromise between computational efficiency and the reliability of the obtained electronic structures of actinide species. 
Furthermore, the actinide semi-core 5s, 5p, and 5d orbitals are far less important for electron correlation than the valence 6s, 6p, and 5f orbitals.\mbox{\cite{real09,tecmer2014,pawel_pccp2015,ptThS}}}
\subsection{EOM-CCSD and CR-EOM-CCSD(T)}
All EOM-CCSD/CR-EOM-CCSD(T)~\cite{bartlett-eom,cr-eomccsd} calculations were carried out in the \textsc{NWChem} (version 6.8) software package~\cite{nwchem,nwchem_web1} using the tensor contraction engine~\cite{tce_1,tce_2,tce_3} and imposing $C_{2v}$ (ThO, ThS, CUO, \ce{ThO_2} and \ce{NUO^+}) and $D_{2h}$ (\ce{UO_2^{2+}}, \ce{NUN}, and \ce{PaO_2^+}) point group symmetry, respectively. 
Due to technical difficulties, we were not able to optimize the excited states in all symmetries for \ce{NUO^+} using \textsc{NWChem}.
Thus, the missing EOM-CCSD excitation energies were calculated (with the same computational setup as above) using the \textsc{Molpro2012} software package.~\cite{molpro2012,molpro-wires} 
\subsection{EOM-pCCD+S, EOM-pCCD-CCS, and EOM-pCCD-LCCSD}
All pCCD,~\cite{limacher_2013,oo-ap1rog,tamar-pcc} EOM-pCCD+S,~\cite{eom-pccd,eom-pccd-erratum} EOM-pCCD-CCS, and EOM-pCCD-LCCSD~\cite{eom-pccd-lccsd,ola-book-chapter-2019} calculations were performed using our locally developed~\textsc{PIERNIK}~\cite{piernik100} software package. 
All EOM-pCCD-based calculations were carried out using $C_1$ point group symmetry to obtain a completely smooth potential energy surface for the pCCD-LCCSD reference calculation (\textit{vide infra}). 
Additional cross-check calculations were performed imposing $D_{2h}$ and ${C_{2v}}$ point group symmetry.
The corresponding electronic spectra are similar to those obtained by relaxing the symmetry constraint and are thus reported in the ESI\dag.
Specifically, the presence of symmetry does significantly influence the EOM-pCCD-LCCSD excitation energies where differences amount to (at most) 0.1 eV.
However, we should note that for stretched inter-atomic distances EOM-pCCD-LCCSD does not provide smooth potential energy surfaces for some excited states (see sections S9 and S10 of the ESI\dag).
This unsmoothness can be cured by relaxing the symmetry constraint.
Thus, a variational orbital optimization procedure within pCCD~\cite{oo-ap1rog} was applied in all pCCD-based calculations for the ground state potential energy curve of \ce{UO_2^{2+}}, \ce{NUN}, \ce{ThO_2}, and \ce{PaO_2^+}, imposing $C_1$ point group symmetry.
Finally, we should note that the orbital optimization induces localization of molecular orbitals in all investigated systems.
Nonetheless, these optimized orbitals can still be identified as $\sigma$-, $\pi$-, $\delta$-, and $\phi$-type orbitals, which facilitates the identification of excited states.

\subsection{Fitting procedure}
All potential energy curves were obtained from a polynomial fit of 8-th order using fitting scripts available in the \textsc{Piernik} software package.
The corresponding spectroscopic constants (equilibrium bond length ($r_{\rm e}$) and harmonic vibrational frequency ($\omega_{\rm e}$)) were calculated based on those fitted potential energy curves. 
Specifically, the harmonic vibrational frequencies ($\omega_{\rm e}$) were determined numerically using the five-point finite difference stencil~\cite{Abramowitz} and the following averaged masses: uranium: 238.0508, thorium: 232.0381, protactinium: 231.0359, oxygen: 15.9949, sulfur: 31.9721, and nitrogen: 14.0031.~\cite{spectro-data-2005} 

\begin{table*}[ht]	
\caption{Spectroscopic parameters of the ground and some low-lying singlet excited states of ThO and ThS obtained from various EOM-CC models. \re\ is the equilibrium bond length, \we\ the harmonic vibrational frequency, \Ta\ the adiabatic excitation energy, and \Tv\ the vertical excitation energy, respectively.
\Tv\ has been calculated with respect to the equilibrium bond distance \re\ of the corresponding ground state obtained for each CC method. The differences with respect to CCSD/CR-EOM-CCSD(T) are given in parentheses. The total electronic energies, excitation energies, and excited states PES are summarized in the ESI\dag.} 
\centering
\label{tbl:thoths}
{\footnotesize
\begin{tabular}{l|cccc|cccc}
    & \multicolumn{4}{c|}{ThO} & \multicolumn{4}{c}{ThS} \\ \cline{2-9}
	Method & \re\ [\AA]  & \we\ [cm$^{-1}$] & \Ta\ [eV] & \Tv [eV] & \re\ [\AA]  & \we\ [cm$^{-1}$] & \Ta\ [eV] & \Tv [eV]    \\\hline \hline 
    
	& \multicolumn{4}{c|}{X$^1\Sigma$} & \multicolumn{4}{c}{X$^1\Sigma$}\\ \hline 
	pCCD                   & 1.850($-$0.001)    & 918($+$36) & - & - & 2.386($-$0.029) & 497($+$5) & -  & - \\ 
	pCCD-CCS               & 1.851($\pm$0.000)  & 900($+$18) & - & - & 2.386($+$0.029) & 494($+$2) & -  & - \\ 
	pCCD-LCCSD             & 1.871($+$0.020)    & 834($-$48) & - & - & 2.387($+$0.030) & 436($-$56)& -  & - \\ 
	CCSD                   & 1.851              & 882        & - & - & 2.357 & 492 & -     & - \\ 
    &&&&&&&\\
	& \multicolumn{4}{c|}{$1^1\Delta$ $(\sigma\rightarrow\delta)$}  & \multicolumn{4}{c}{$1^1\Delta(\sigma\rightarrow\delta)$}  \\ \hline 
	EOM-pCCD+S             & 1.860($-$0.002) & 875($+$21) & 2.03($+$0.76) & 2.03($+$0.76) & 2.410($+$0.029) & 486($+$9 ) & 1.88($+$0.70) & 1.89($+$0.70) \\ 
	EOM-pCCD-CCS           & 1.859($-$0.003) & 884($+$30) & 2.04($+$0.77) & 2.04($+$0.77) & 2.410($+$0.029) & 478($+$1)  & 1.90($+$0.72) & 1.90($+$0.71) \\ 
	EOM-pCCD-LCCSD         & 1.881($+$0.019) & 819($-$35) & 1.56($+$0.29) & 1.56($+$0.29) & 2.405($+$0.024) & 431($-$46) & 1.56($+$0.38) & 1.56($+$0.37) \\ 
	EOM-CCSD               & 1.864($+$0.002) & 846($-$8)  & 1.37($+$0.10) & 1.38($+$0.11) & 2.384($+$0.003) & 474($-$3)  & 1.29($+$0.11) & 1.30($+$0.11) \\ 
	CR-EOM-CCSD(T)            & 1.862 & 854 & 1.27 & 1.27  &2.381 & 477 & 1.18 & 1.19 \\ 
    &&&&&&&\\
	& \multicolumn{4}{c|}{$1^1\Pi$ $(\sigma\rightarrow\pi)$}    & \multicolumn{4}{c}{$1^1\Pi(\sigma\rightarrow\pi)$}  \\ \hline
	EOM-pCCD+S             & 1.881($-$0.004)   & 854($+$13) & 2.82($+$0.92) & 2.84($+$0.92)& 2.441($+$0.028) & 467($+$10) & 2.32($+$0.84) & 2.35($+$0.83)\\ 
	EOM-pCCD-CCS           & 1.881($-$0.004)   & 871($+$30) & 2.82($+$0.92) & 2.84($+$0.92)& 2.441($+$0.028) & 461($+$4)  & 2.33($+$0.85) & 2.36($+$0.84) \\
	EOM-pCCD-LCCSD         & 1.900(+0.015)     & 825($-$16) & 2.24($+$0.34) & 2.26($+$0.34)& 2.434($+$0.021) & 420($-$37) & 1.95($+$0.47) & 1.97($+$0.45)\\ 
	EOM-CCSD               & 1.885($\pm$0.000) & 842($+$1)  & 2.05($+$0.15) & 2.07($+$0.15)& 2.415($+$0.002) & 461($+$4)  & 1.65($+$0.17) & 1.69($+$0.17) \\ 
	CR-EOM-CCSD(T)             & 1.885 & 841 & 1.90 & 1.92 & 2.413 & 457 & 1.48 & 1.52  \\ 
    &&&&&&&\\
	& \multicolumn{4}{c|}{$1^1\Sigma$ $(\sigma\rightarrow\sigma)$}   & 	\multicolumn{4}{c}{$1^1\Sigma(\sigma\rightarrow\sigma)$}  \\ \hline 
	EOM-pCCD+S             & 1.892($+$0.007) & 885($+$49) & 2.57($+$0.34) & 2.61($+$0.36)  & 2.455($+$0.038)  & 460($\pm$0) & 1.94($+$0.07) & 2.00($+$0.09)\\
	EOM-pCCD-CCS           & 1.891($+$0.006) & 848($+$12) & 2.57($+$0.34) & 2.60($+$0.35)  & 2.454($+$0.037)  & 461($+$1)   & 1.94($+$0.07) & 2.00($+$0.09)  \\ 
	EOM-pCCD-LCCSD         & 1.908($+$0.023) & 786($-$50) & 2.27($+$0.04) & 2.30($+$0.05)  & 2.462($+$0.045)  & 355($-$105) & 1.88($+$0.01) & 1.92($+$0.01) \\
	EOM-CCSD               & 1.884($-$0.001) & 839($+$3)  & 2.36($+$0.13) & 2.38($+$0.13)  & 2.417($\pm$0.000)& 456($-$4)   & 2.02($+$0.15) & 2.07($+$0.16)   \\ 
	CR-EOM-CCSD(T)             & 1.885 & 836 & 2.23 & 2.25 & 2.417            & 460 & 1.87 & 1.91\\ 
    &&&&&&&\\
	& \multicolumn{4}{c|}{$2^1\Delta$ $(\sigma\rightarrow\delta)$} & \multicolumn{4}{c}{$2 ^1\Delta(\sigma\rightarrow\delta) $ } \\ \hline
	EOM-pCCD+S             & 1.878($-$0.004)   & 838($+$10) & 3.28($+$0.28) & 3.29($+$0.27)  & 2.442($+$0.039) & 463($-$3)   & 2.76($+$0.13) & 2.80($+$0.14)     \\ 
	EOM-pCCD-CCS           & 1.880($-$0.002)   & 856($+$28) & 3.28($+$0.28) & 3.30($+$0.28)  & 2.442($+$0.039) & 457($-$9)   & 2.76($+$0.13) & 2.80($+$0.14)     \\ 
	EOM-pCCD-LCCSD         & 1.909($+$0.027)   & 771($-$57) & 2.68($-$0.32) & 2.70($-$0.32)  & 2.471($+$0.058) & 318($-$148) & 2.17($-$0.46) & 2.22($-$0.44)   \\
	EOM-CCSD               & 1.882($\pm$0.000) & 855($+$27) & 4.51($+$1.51) & 4.54($+$1.52)  & 2.412($+$0.009) & 460($-$6)   & 4.24($+$1.61) & 4.27($+$1.61)     \\ 
	CR-EOM-CCSD(T)             & 1.882 & 828 & 3.00 & 3.02 & 2.403 &  466 & 2.63 & 2.66   \\ 
\hline \hline
\end{tabular}
}
\end{table*}
\section{Adiabatic electronic spectra}\label{sec:results}
In the following, we first scrutinize the lowest-lying excited states of the ThO and ThS diatomic molecules, followed by the electronic spectra of the \ce{UO2^2+} molecule and its isoelectronic series.

\subsection{ThO and ThS}\label{sec:tho-and-ths}

\begin{figure*}[h!]
\includegraphics[width=1.0\textwidth]{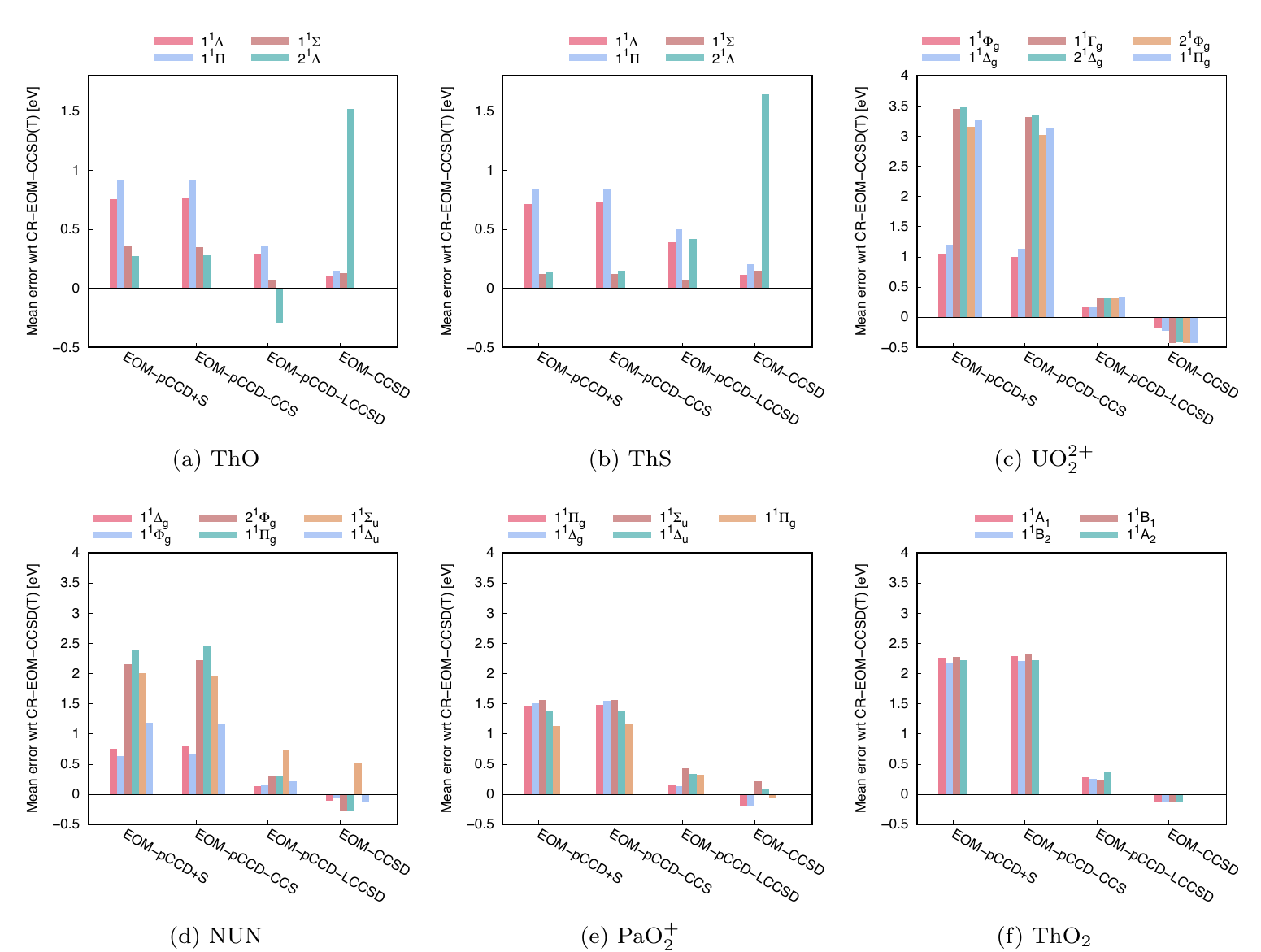}
\caption{Mean errors in excitation energies with respect to CR-EOM-CCSD(T) reference data calculated from eq.~\eqref{eq:me} for all investigate actinide-containing compounds. All error measures are summarized in the ESI\dag.}
\label{fig:me_surface}
\end{figure*}

ThO and ThS can be considered as the simplest actinide representatives.
Despite their simplicity, both compounds are of considerably growing importance in the search of the electron electric dipole moment (eEDM).~\cite{heaven:barker:antonov,wang:le:steimle:heaven,ACME_science,Sulfur_Heaven_2014,ptThS} 
While the electronic spectrum of ThO has been investigated using many quantum chemistry methods,~\cite{Marian_tho_1988,theo_casscf_paulovic,PhysRevA.78.010502,edm_fleig_nayak2014,Skripnikov-ThO-2015,Skripnikov-ThO-2016,Denis-ThO-2016,ptThS} the electronic excitations of ThS were studied only recently.~\cite{ThS_liang_andrews_2002,ptThS} 
Hence, both molecules are good test systems to assess the performance of various EOM-pCCD-based approaches, which to the best of our knowledge, have never been applied to such systems before. 

The left hand side of Table~\ref{tbl:thoths} lists all ground and excited states spectroscopic parameters of ThO obtained from different coupled cluster methods.
The numerical values for the total electronic energies as well as excitation energies are summarized in the ESI\dag.
The ground state equilibrium bond lengths of 1.85-1.87~\AA~are in good agreement with the experimental value of 1.84~\AA.~\cite{exp_goncharov1,exp_dewberry} 
{For comparison, experimental spectroscopic parameters are presented in Table S23 of the ESI\dag.}  
Furthermore, the vibrational frequencies (between 830 and 920 cm$^{-1}$) are in reasonable agreement with the experimental spectrum ranging from 879 to 896 cm$^{-1}$.~\cite{exp_edvinsson,exp_andrews_jacs,exp_kushto}
The optimal bond lengths of the electronic ground state obtained from both pCCD and pCCD-CCS are very close to the value of 1.851~\AA~predicted by CCSD. 
Adding an LCCSD correction on top of pCCD considerably elongates \re\ by 0.02~\AA. 
Furthermore, pCCD and pCCD-CCS overestimate the vibrational frequency by only 20--40 cm$^{-1}$, while pCCD-LCCSD underestimates \we\ by about 50 cm$^{-1}$.

In the lowest part of the electronic spectrum of ThO, the electronic excited states feature electron transfer from the $\sigma$ highest occupied molecular orbital (HOMO) (with leading contributions from the Th 7s orbital) to the unoccupied $\delta$, $\pi$, and $\sigma$ molecular orbitals (with dominant contributions from the Th 6d and 7p orbitals, respectively). 
Specifically, electrons are transferred to the Th 6d$_\delta$ orbital in the $1^1\Delta$ state, to the Th 7p$_\sigma$ orbital in the $1^1\Sigma$ state, and to the mixed Th 6d$_\pi$/7p$_\pi$ orbitals in the $1^1\Pi$ state, while the $2^1\Delta$ state features a double electron transfer to the Th 6d$_\delta$ orbital. 
This is in line with previous theoretical studies of excitation energies in ThO.~\cite{theo_casscf_paulovic,ptThS}
Furthermore, all EOM-CC-based methods give similar values for vertical and adiabatic excitation energies, suggesting only minor importance of structure relaxation upon electron excitation. 
The small deviations in \re\ of all targeted excited states as well as the orbital contributions to the excitation energies, which are almost entirely dominated by atomic orbitals of Th, confirm this minor dependence of electronic excitations on the molecular structure.
In general, all EOM-pCCD-based methods provide spectroscopic constants and excitation energies that are qualitatively correct compared to CR-EOM-CCSD(T) reference data.
Specifically, EOM-pCCD+S and EOM-pCCD-CCS overestimate excited state vibrational frequencies by 10--70 cm$^{-1}$, while EOM-pCCD-LCCSD underestimates them by 10--60 cm$^{-1}$. 
EOM-pCCD-LCCSD provides, however, excitation energies that are much closer to CR-EOM-CCSD(T) results with differences amounting to at most 0.3 eV. 
One exception is the doubly excited 2 $^1\Delta$, for which EOM-pCCD-based methods should be considered as the reference.~\cite{eom-pccd-lccsd,pawel-yb2}

The right hand side of Table~\ref{tbl:thoths} collects all spectroscopic constants of the ground and four low-lying excited states for ThS. 
The major differences in spectroscopic constants with respect to ThO are the longer equilibrium bond lengths and smaller electron excitation energies.
We should emphasize that the triplet excited states in ThS, which are not considered in this work, are observed at much lower energy ranges than found in the ThO molecule.
Similar to ThO, the electronic excitations in ThS occur from the doubly occupied $\sigma$ HOMO to the unoccupied Th 6d and 7p orbitals. 
The CCSD ground state equilibrium bond distance of 2.357~\AA~and vibrational frequency of 492 cm$^{-1}$ fall in line with previous theoretical and experimental investigations.~\cite{ThS_liang_andrews_2002,ThS_Barlett_Antonov_Heaven2013,Sulfur_Heaven_2014,ptThS}  
Furthermore, all excited states in the ThS molecule are only marginally effected by changes in the molecular structure (cf. Table~\ref{tbl:thoths}). 
Despite the similar character of excited states present in ThO and ThS, the performance of EOM-pCCD-based methods is slightly worse for ThS compared to ThO. 
Specifically, the differences between EOM-pCCD-based methods with respect to CR-EOM-CCSD(T) increase to 0.5 eV. 
An exception is again the doubly excited 2$^1\Delta$ state. 
Finally, we should note that the triple correction within the CR-EOM-CCSD(T) approach has only a minor effect on spectroscopic parameters in both ThO and ThS.   

Figures~\ref{fig:me_surface}(a) and (b) display the mean error (ME) in excitation energies along each PES with respect to CR-EOM-CCSD(T) for ThO and ThS, respectively, calculated from
\begin{equation}\label{eq:me}
    \mathrm{ME}=\sum\limits_{r_{\rm AcL}}\Delta \omega_{r_{\rm AcL}}/N,
\end{equation}
for all actinide--ligand (Ac--L) distances $r_{\rm AcL}$ along each PES, where $N$ is the total number of points along the PES and $\Delta \omega_{r_{\rm AcL}} = \omega^{\rm CC}_{r_{\rm AcL}}-\omega^{\rm ref}_{r_{\rm AcL}}$.
We have determined additional error measures for both excitation energies $\omega$ and total electronic energies $E$ (along the PES), which are summarized in section S7 of the ESI\dag.
These include the non-parallelity error NPE,
\begin{equation}
    \mathrm{NPE} = \max\limits_{r_{\rm AcL}}(|\Delta E_{r_{\rm AcL}}|) - \min\limits_{r_{\rm AcL}} (|\Delta E_{r_{\rm AcL}}|),
\end{equation}
the maximum absolute error MAE,
\begin{equation}
    \mathrm{MAE}(x) =\max\limits_{r_{\rm AcL}}(|\Delta x_{r_{\rm AcL}}|),
\end{equation}
where $x$ is either the total electronic energy at a given interatomic distance $E_{r_{\rm AcL}}$ or the (vertical) excitation energy $\omega$ at a given ${r_{\rm AcL}}$, and the root mean error RME,
\begin{equation}
    \mathrm{RME}=\sqrt{\sum\limits_{r_{\rm AcL}}\Delta x_{r_{\rm AcL}}^2/N},
\end{equation}
similarly evaluated for both total electronic energies at a given interatomic distance $E_{r_{\rm AcL}}$ and (vertical) excitation energies $\omega$ at a given ${r_{\rm AcL}}$.
In general, the simple EOM-pCCD+S and EOM-pCCD-CCS methods have large error measures of about 0.5 to 1.0 eV, while EOM-pCCD-LCCSD and EOM-CCSD have negligibly small MAE, ME, and RME.
In general, EOM-pCCD-LCCSD yields mean errors that are slightly larger than those observed for EOM-CCSD.
However, both methods predict excitation energies that differ by at most 0.5 eV from the CR-EOM-CCSD(T) reference data.
One exception is the doubly-excited 2$^1\Delta$ state, where EOM-CCSD completely fails and predicts errors of more than 1.5 eV.
Finally, we should note that the error measures for the total electronic energies are significantly larger as the ground and excited state potential energy surfaces are shifted in energy with respect to the CR-EOM-CCSD(T) reference curve.
Hence, the corresponding data is presented in Table S16 of the ESI\dag{} for reasons of completeness.

\subsection{The \ce{UO_2^{2+}} isoelectronic series}\label{sec:isoelectronic}
The uranyl cation (\ce{UO_2^{2+}}) is often considered as the most prevalent building block of many uranium-containing compounds.~\cite{denning2007,baker2012}
Quantum chemical studies on the ground and excited-state electronic structures of the uranyl fragment have been an active field of research over the past years.~\cite{Wadt1981,dyall_1999,uranyl_de_jong_99,uranyl_pitzer_99,kaltsoyannis,matsika_2001,real07,real09,pawel1,fen_wei,pawel2,pawel3,pawel_saldien,tecmer2014,gomes2015applied,gomes_crystal,tecmer-song2016}
As a result, its electronic structure is now well-established, featuring a singlet electronic ground state, a linear geometry, and characteristic vibrational and electronic spectra.~\cite{tecmer-song2016}
Hence, combined with its isoelectronic actinide-containing analogs, \ce{UO_2^{2+}} represents a stringent test case for assessing the accuracy of new approximate quantum chemistry approaches.
By comparing ground and excited-state electronic structures across the uranyl isoelectronic series, we obtain valuable information about the validity and reliability of a given electronic structure method. 
In this work, the uranyl isoelectronic set is composed of the \ce{UO_2^{2+}}, NUN, \ce{PaO_2^+}, \ce{ThO_2}, \ce{NUO^+}, and CUO molecules. 
Specifically, we focus on vertical excitation energies in all of the above mentioned species as well as on the adiabatic excitation energies of the \ce{UO_2^{2+}}, NUN, \ce{PaO_2^+}, and \ce{ThO_2} moieties.
We should note that the electronic structure of \ce{ThO2} is completely different from the other investigated triatomic molecules isoelectronic to the uranyl cation.
It features a bent geometry and a peculiar electronic structure, where the Th 5f orbitals do not participate neither in bonding nor in electronic excitations. 
These differences originate from the relative energetic ordering of the 5f and 6d orbitals in the thorium and uranium atoms, respectively.  
Specifically, for the \ce{ThO2} molecule, the thorium 6d orbitals are lower in energy and dominate in back-bonding, while in the \ce{UO2^2+} cation the uranium 5f orbitals are lower in energy and thus outweigh the oxygen atoms in the back-bonding process.~\cite{Wadt1981}

\begin{sidewaystable*}[tp]
	\centering
	\scriptsize
	\caption{Spectroscopic parameters of the ground and low-lying singlet excited states of \ce{UO2^2+}, NUN, \ce{PaO2+}, and \ce{ThO2} from various EOM-CC models. \re\ is the equilibrium bond length, \we\ the harmonic vibrational frequency, \Ta\ the adiabatic excitation energy, and \Tv\ the vertical excitation energy, respectively.
    \Tv\ has been calculated with respect to the equilibrium bond distance \re\ of the corresponding ground state obtained for each CC method.
    The differences with respect to CCSD/CR-EOM-CCSD(T) are given in parentheses. The total electronic energies, excitation energies, and excited states PES are summarized in the ESI\dag.} 
	\label{tbl:adiabatic-spectrum}
	\scalebox{0.8}{
		\begin{tabular}{l|cccc|cccc|cccc|cccc}
			& \multicolumn{4}{c|}{\ce{UO2^2+}} 
			& \multicolumn{4}{c|}{NUN} 
			& \multicolumn{4}{c|}{\ce{PaO2+}}
			& \multicolumn{4}{c}{\ce{ThO2} ($\measuredangle$ (O--Th--O) = 120$\degree$)} \\ \cline{2-17}
			
			Method & \re\ [\AA]  & \we\ [cm$^{-1}$] & \Ta\ [eV] & \Tv [eV] 
			& \re\ [\AA]  & \we\ [cm$^{-1}$] & \Ta\ [eV] & \Tv [eV]  
			& \re\ [\AA]  & \we\ [cm$^{-1}$] & \Ta\ [eV] & \Tv [eV] 
			& \re\ [\AA]  & \we\ [cm$^{-1}$] & \Ta\ [eV] & \Tv [eV]    \\\hline \hline

			& \multicolumn{4}{c|}{X$^1\Sigma_g$} & \multicolumn{4}{c|}{X$^1\Sigma_g$}&	\multicolumn{4}{c|}{X$^1\Sigma_g$} & \multicolumn{4}{c}{X$^11A_1$}\\ \hline 
			
			pCCD             & 1.670($-$0.020)  &1139($+$48)  &  &  & 1.711($-$0.016)  & 1176($+$63) &  &  & 1.763($-$0.007) &972($+$7)   &  &   & 1.917($-$0.001)   & 805($+$1)  &      &  \\   
			pCCD-CCS         & 1.670($-$0.020)  &1136($+$45)  &  &  & 1.711($-$0.016)  & 1161($+$48) &  &  & 1.763($-$0.007) &970($+$5)   &  &   & 1.918($\pm$0.000) & 807($+$3)  &      &  \\   
			pCCD-LCCSD       & 1.717($+$0.027)  & 976($-$115) &  &  & 1.748($+$0.021)  & 960($-$153) &  &  & 1.789($+$0.019) & 895($-$70) &  &   & 1.934($+$0.016)   & 773($-$31) &      &  \\   
			CCSD             & 1.690            & 1091        &  &  & 1.727            &  1113       &  &  & 1.770 &   965  &  &  & 1.918  & 804 &                                      &   \\   
			&&&&&&&&&&&&\\
			& \multicolumn{4}{c|}{$1^1\Phi_g(\sigma_u\rightarrow\phi_u)$} & \multicolumn{4}{c|}{$1^1\Phi_g(\sigma_u\rightarrow\phi_u)$}  &\multicolumn{4}{c|}{$1^1\Phi_g(\sigma_u\rightarrow\phi_u)$}   &\multicolumn{4}{c}{$1 ^1A_1(a_1 \rightarrow a_1)$} \\ \hline 
			
			EOM-pCCD+S        & 1.708($-$0.047)  &1028($+$91)  &4.80($+$1.04)  &4.90($+$0.91)& 1.747($-$0.038) & 1104($+$141) & 3.40($+$0.71) & 3.48($+$0.62) & 1.789($-$0.059) & 962($+$148) &6.24($+$1.56) &6.28($+$1.34) & 1.961($-$0.021) & 744($+$29) & 5.49($+$2.30) & 5.55($+$2.24) \\
			EOM-pCCD-CCS      & 1.703($-$0.052)  &1057($+$120) &4.83($+$1.07)  &4.91($+$0.92)& 1.746($-$0.039) & 1097($+$134) & 3.42($+$0.73) & 3.50($+$0.64)   & 1.783($-$0.065) & 925($+$111) &6.27($+$1.59) &6.30($+$1.36) & 1.964($-$0.018) & 742($+$27) & 5.50($+$2.31) & 5.57($+$2.26) \\
			EOM-pCCD-LCCSD    & 1.786($+$0.031)  & 837($-$100) &3.66($-$0.10)  &3.87($-$0.12)& 1.811($+$0.026) & 893($-$70)   & 2.66($-$0.03) & 2.83($-$0.03)   & 1.863($+$0.015) & 807($-$7)   &4.66($-$0.22) &4.88($-$0.06) & 1.997($+$0.015) & 711($-$4)  & 3.35($+$0.16) & 3.46($+$0.15) \\
			EOM-CCSD          &1.762($+$0.007)   & 924($-$13)  &3.57($-$0.19)  &3.86($-$0.13)& 1.789($+$0.004) & 952($-$11)   & 2.64($-$0.05) & 2.83($-$0.03)  & 1.858($+$0.010) & 804($-$8)   &4.47($-$0.21) &4.79($-$0.15) & 1.984($+$0.002) & 716($+$1)  & 3.06($-$0.13) & 3.19($-$0.12) \\
			CR-EOM-CCSD(T)    &1.755             &937          &3.76           & 3.99        & 1.785           &  963         & 2.69          & 2.86   & 1.848           & 814         & 4.68         &4.94          &1.982            & 715        & 3.19          & 3.31          \\
			&&&&&&&&&&&&\\
			& \multicolumn{4}{c|}{$1^1\Delta_g(\sigma_u\rightarrow\delta_u)$}   & 	\multicolumn{4}{c|}{$1^1\Delta_g(\sigma_u\rightarrow\delta_u)$}&\multicolumn{4}{c|}{$1^1\Delta_g(\sigma_u\rightarrow\delta_u)$}  &\multicolumn{4}{c}{$1^1B_2(b_2 \rightarrow a_1)$}   \\ \hline     
			EOM-pCCD+S        &1.706($-$0.046)  & 1014($+$81)  &5.34($+$1.18)  &5.42($+$1.05) & 1.745($-$0.035) & 1075($+$100) & 4.07($+$0.83) &4.14($+$0.75) &1.790($-$0.056)  & 904($+$98)  &6.69($+$1.62) &6.72($+$1.41) & 1.920($-$0.043) & 760($+$53) & 4.79($+$2.16) & 4.79($+$2.10) \\
			EOM-pCCD-CCS      & 1.701($-$0.051) & 1049($+$116) &5.37($+$1.21)  &5.43($+$1.06) & 1.744($-$0.036) & 1076($+$101) & 4.09($+$0.85) &4.16($+$0.77)  &1.781($-$0.065)  & 933($+$127) &6.72($+$1.65) &6.74($+$1.43) & 1.925($-$0.038) & 766($+$59) & 4.81($+$2.18) & 4.81($+$2.12) \\ 
			EOM-pCCD-LCCSD    & 1.781($+$0.029) & 840($-$93)   &4.07($-$0.09)  & 4.25($-$0.12)& 1.807($+$0.027) & 865($-$100)  & 3.22($-$0.02) &3.36($-$0.03)   &1.860($+$0.014)  & 803($-$3)   &5.04($-$0.03) &5.25($-$0.06) & 1.975($+$0.012) & 706($-$1)  & 2.78($+$0.15) & 2.83($+$0.14) \\
			EOM-CCSD          &1.760($+$0.008)  & 910($-$23)   &3.96($-$0.20)  &4.22($-$0.15) & 1.789($+$0.009) & 937($-$38)   & 3.14($-$0.10) &3.34($-$0.05)   &1.856($+$0.010)  & 798($-$8)   &4.86($-$0.21) &5.16($-$0.15) & 1.970($+$0.007) & 714($+$7)  & 2.49($-$0.14) & 2.57($-$0.12) \\ 
			CR-EOM-CCSD(T)    &1.752            &933           &4.16           &4.37          & 1.780           & 975          & 3.24          &3.39         &1.846            & 806         &5.07          & 5.31         & 1.963           & 707        & 2.63          & 2.69          \\ 
			&&&&&&&&&&&&\\
			& \multicolumn{4}{c|}{$2^1\Phi_g(\pi_u\rightarrow\delta_u)$}   & \multicolumn{4}{c|}{$2^1\Phi_g(\pi_u\rightarrow\delta_u)$} &\multicolumn{4}{c|}{}    &\multicolumn{4}{c}{$1^1B_1(b_1 \rightarrow a_1)$}   \\ \hline
			EOM-pCCD+S         & 1.762($-$0.028) & 962($+$56) &8.17($+$3.27)  &8.65($+$3.20)& 1.776($-$0.031) & 971($+$2)  & 6.14($+$2.32) & 6.38($+$2.25)  &&&& 
			& 2.011($-$0.013)   & 718($+$15) & 5.81($+$2.32) & 6.11($+$2.26)\\
			EOM-pCCD-CCS       &1.749($-$0.041)  & 943($+$37) &8.32($+$3.42)  &8.69($+$3.24)& 1.772($-$0.035) & 986($+$17) & 6.20($+$2.38) & 6.42($+$2.29)   &&&& & 2.012($-$0.012)   & 715($+$12) & 5.85($+$2.36) & 6.14($+$2.29)\\
			EOM-pCCD-LCCSD    &1.804($+$0.014)  & 941($+$35) &4.87($-$0.03)  &5.29($-$0.16)& 1.822($+$0.015) & 938($-$31) & 3.90($+$0.08) & 4.16($+$0.03)   &&&& & 2.035($+$0.011)   & 690($-$13) & 3.61($+$0.12) & 3.92($+$0.06)\\
			EOM-CCSD          & 1.797($+$0.007) & 902($-$4)  &4.46($-$0.44)  &5.08($-$0.37)& 1.810($+$0.003) & 946($-$23) & 3.54($-$0.28) & 3.90($-$0.23)   &&&& & 2.024($\pm$0.000) & 706($+$3)  & 3.35($-$0.14) & 3.71($-$0.14)\\
			CR-EOM-CCSD(T)   & 1.790           & 906        &4.90           &5.45         & 1.807           & 969        & 3.82          & 4.13            &&&&       & 2.024             & 703        & 3.49          & 3.85         \\
			&&&&&&&&&&&&\\
			& \multicolumn{4}{c|}{$1^1\Pi_g(\pi_u\rightarrow\delta_u)$} & \multicolumn{4}{c|}{$1^1\Pi_g(\pi_u\rightarrow\delta_u)$} &\multicolumn{4}{c|}{$1^1\Pi_g(\sigma_u\rightarrow\pi_u)$} &\multicolumn{4}{c}{$1^1A_2(b_2 \rightarrow b_1) $ } \\ \hline
			EOM-pCCD+S        &1.763($-$0.027)  &963($+$58) &8.35($+$3.36)  &8.84($+$3.31)& 1.780($-$0.025) & 949($-$10)  & 6.51($+$2.54) & 6.80($+$2.52) & 1.825($-$0.040) & 844($+$21)  & 7.14($+$1.30)  &7.31($+$1.09) & 1.932($-$0.055) & 679($+$84)  & 6.18($+$2.17) & 6.19($+$2.07)\\
			EOM-pCCD-CCS       &1.749($-$0.041)  &952($+$47) &8.50($+$3.51)  &8.87($+$3.34)& 1.776($-$0.029) & 1016($+$57) & 6.56($+$2.59) & 6.82($+$2.54)  & 1.820($-$0.045) & 790($-$33)  & 7.17($+$1.33)  &7.33($+$1.11) & 1.930($-$0.057) & 709($+$114) & 6.20($+$2.19) & 6.21($+$2.09)\\
			EOM-pCCD-LCCSD    & 1.804($+$0.014) &965($+$61) &4.98($-$0.01)  &5.38($-$0.15)& 1.821($+$0.016) & 935($-$24)  & 4.07($+$0.10) & 4.32($+$0.05)  & 1.890($+$0.025) & 759($-$64)  & 5.97($+$0.13)  &6.33($+$0.11)  & 1.986($-$0.001) & 681($+$86)  & 4.20($+$0.19) & 4.28($+$0.06)\\
			EOM-CCSD           & 1.796($-$0.006) &904($-$1)  &4.55($-$0.44)  &5.16($-$0.37)& 1.811($+$0.006) & 946($-$13)  & 3.68($-$0.29) & 4.04($-$0.24)  & 1.884($+$0.019) & 757($-$66)  & 5.73($-$0.11)  &6.22($\pm$0.00) & 1.994($+$0.007) & 589($-$6)   & 3.89($-$0.12) & 4.02($-$0.10)\\
			CR-EOM-CCSD(T)    & 1.790           &905        &4.99           &5.53               & 1.805           & 959         & 3.97          & 4.28            & 1.865           & 823         & 5.84           & 6.22        & 1.987           & 595         & 4.01          & 4.12         \\
			&&&&&&&&&&&&\\
			& \multicolumn{4}{c|}{$2^1\Delta_g(\pi_u\rightarrow\phi_u)$}  &\multicolumn{4}{c|}{$1^1\Sigma_u(\sigma_u\rightarrow\sigma_g)$}  &\multicolumn{4}{c|}{$1^1\Sigma_u(\sigma_u\rightarrow\sigma_g)$}    \\ \hline
			EOM-pCCD+S        & 1.767($-$0.040) &932($+$31)  & 8.95($+$3.68) &9.60($+$3.57)& 1.748($-$0.019) & 1079($+$66) & 4.91($+$2.09) & 5.00($+$2.09) & 1.784($-$0.023) & 885($+$15) &6.23($+$1.57) &6.25($+$1.53)   \\ 
			EOM-pCCD-CCS      & 1.752($-$0.055) &1000($+$99) & 9.08($+$3.81) &9.61($+$3.58)& 1.749($-$0.018) & 1080($+$67) & 4.87($+$2.05) & 4.97($+$2.06) & 1.783($-$0.024) & 900($+$30) &6.23($+$1.57) &6.25($+$1.53)   \\ 
			EOM-pCCD-LCCSD    & 1.822($+$0.015) &877($-$24)  & 5.24($-$0.03) &5.81($-$0.22)& 1.778($+$0.011) & 973($-$40)  & 3.39($+$0.57) & 3.43($+$0.52)  & 1.825($+$0.018) & 813($-$57) &4.99($+$0.33) &5.04($+$0.32)   \\
			EOM-CCSD          & 1.811($+$0.004) &915($+$14)  & 4.83($-$0.44) &5.64($-$0.39)& 1.759($-$0.008) & 1032($+$19) & 3.31($+$0.49) & 3.37($+$0.46) & 1.812($+$0.005) & 850($-$20) &4.97($+$0.31) &5.06($+$0.34) \\ 
			CR-EOM-CCSD(T)     & 1.807           &901         & 5.27          &6.03                & 1.767           & 1013        & 2.82          & 2.91          & 1.807           & 870        &4.66          &4.72                \\  
			&&&&&&&&&&&&\\
			& \multicolumn{4}{c|}{$1^1\Gamma_g(\pi_u\rightarrow\phi_u)$}  & \multicolumn{4}{c|}{$1^1\Delta_u(\sigma_u\rightarrow\delta_g)$} & \multicolumn{4}{c|}{$1^1\Delta_u(\sigma_u\rightarrow\delta_g)$} & \multicolumn{4}{c}{}  \\ \hline
			EOM-pCCD+S        & 1.778($-$0.028) &969($+$74) &8.63($+$3.60)  &9.31($+$3.56)& 1.734($-$0.002) & 1125($-$17) & 4.86($+$1.28) & 4.89($+$1.30) & 1.775($-$0.035) &873($+$26)  & 6.23($+$1.37) & 6.24($+$1.31)&&&\\
			EOM-pCCD-CCS      & 1.767($-$0.039) &915($+$20) &8.81($+$3.78)  &9.34($+$3.59)& 1.734($-$0.002) & 1115($-$27) & 4.84($+$1.26) & 4.87($+$1.28)& 1.772($-$0.038) &929($+$82)  & 6.23($+$1.37) & 6.24($+$1.31)&&&\\
			EOM-pCCD-LCCSD    & 1.818($+$0.012) &921($+$26) &4.99($-$0.04)  &5.56($-$0.19)& 1.784($+$0.048) & 901($-$241) & 3.76($+$0.18) & 3.81($+$0.22)& 1.830($+$0.020) &779($-$68)  & 5.09($+$0.23) & 5.15($+$0.22)&&&\\
			EOM-CCSD           & 1.812($+$0.006) &889($-$6)  &4.57($-$0.46)  &5.38($-$0.37)& 1.777($+$0.041) & 865($-$277) & 3.63($+$0.05) & 3.73($+$0.14) & 1.816($+$0.006) &786($-$61)  & 4.95($+$0.09) & 5.04($+$0.11)&&&\\
			CR-EOM-CCSD(T)   & 1.806           &895        &5.03           &5.75           & 1.736           & 1142        & 3.58          &3.59          & 1.810           & 847        & 4.86          & 4.93     &&&\\ 
			\hline \hline
		\end{tabular}
	}
\end{sidewaystable*}

 
Before scrutinizing the electronic spectra and the resulting spectroscopic parameters for the uranyl isoelectronic series, we start our discussion with their corresponding ground state properties. 
As shown in Table~\ref{tbl:adiabatic-spectrum}, the CCSD ground state bond lengths increase from 1.690~\AA~in \ce{UO2^{2+}}, to 1.727~\AA~in NUN, to 1.770~\AA~in \ce{PaO2+}, to 1.918~\AA~in \ce{ThO2}, while the CCSD vibrational frequencies decrease in a similar order (the largest value of 1113 cm$^{-1}$ observed for NUN is reduced to 1091 cm$^{-1}$ in \ce{UO2^{2+}}, to 965 cm$^{-1}$ in \ce{PaO2^+}, to 804 cm$^{-1}$ in \ce{ThO2}).
These trends are, to a large extent, reproduced by our simplified pCCD-based models.
Specifically, the pCCD and pCCD-CCS methods underestimate the optimal bond-lengths and overestimate vibrational frequencies. 
The opposite is true for pCCD-LCCSD. 
Nonetheless, the overall accuracy of the investigated pCCD-based models is quite satisfactory compared to CCSD and improves considerably when moving to \ce{PaO2+} and \ce{ThO2} (cf. Table~\ref{tbl:adiabatic-spectrum}). 
From the data available in the literature,~\cite{xuy,pierloot05,real09,nun,kovacs2011,pawel2,kovacs-chem-rev-2015,tecmer-song2016} we can estimate that the ranges for the optimal bond length are 1.67--1.72~\AA~for \ce{UO2^{2+}}, 1.71--1.76~\AA~for NUN, 1.71--1.78~\AA~for \ce{PaO2^+}, and 1.88--1.93~\AA{} for \ce{ThO2}, respectively. 
The characteristic symmetric vibrational frequencies should oscillate around 930--1150 cm$^{-1}$ for \ce{UO2^{2+}}, around 1050 cm$^{-1}$ for NUN, around 900--1000 cm$^{-1}$ for \ce{PaO2^{+}}, and around 800 cm$^{-1}$ for \ce{ThO2}, respectively {(see also Table S23 of the ESI\dag)}. 
Thus, our CCSD and pCCD-based bond lengths and vibrational frequencies fall in line with previous theoretical and experimental findings~\cite{kovacs-chem-rev-2015}.

\begin{figure*}[h!]
\includegraphics[width=1.0\textwidth]{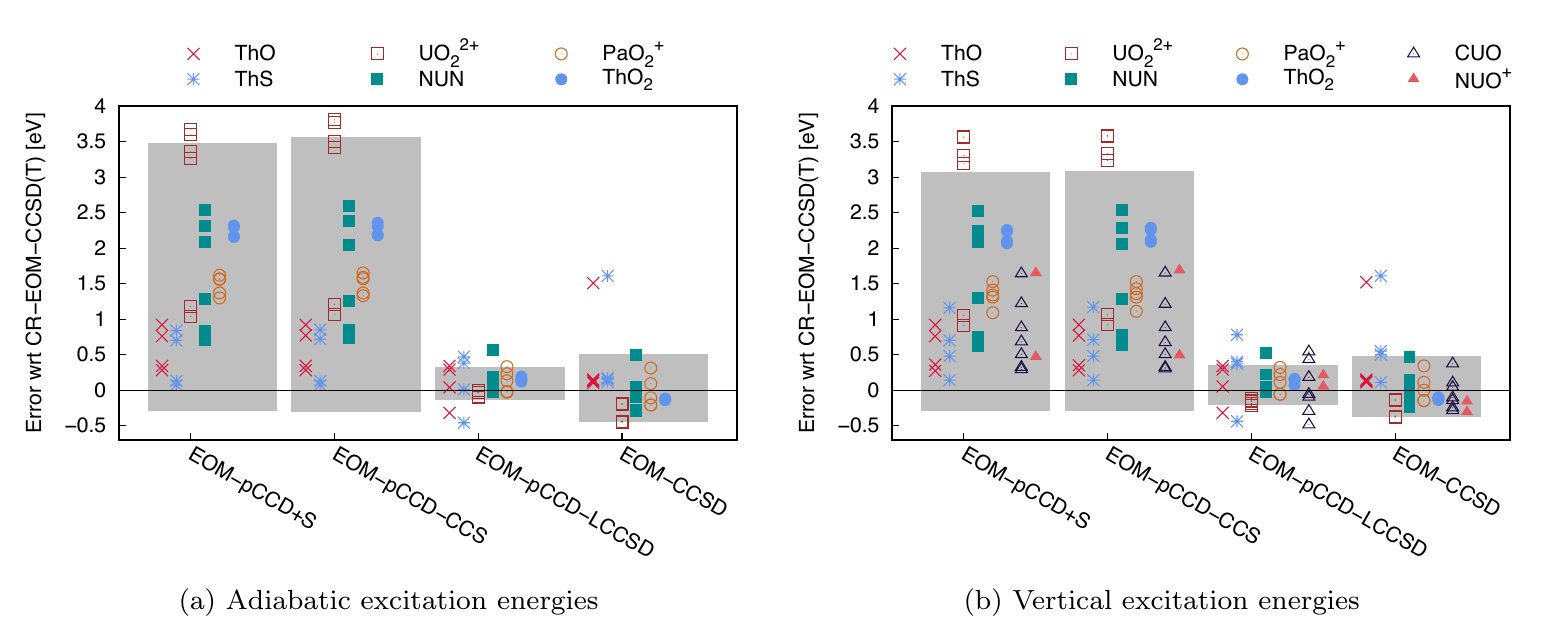}
\caption{Errors in excitation energies with respect to CR-EOM-CCSD(T) reference data for all investigate actinide-containing compounds. (a) Errors in adiabatic excitation energies including the standard deviation (grey box). (b) Errors in vertical excitation energies including the standard deviation (grey box).}
\label{fig:errors}
\end{figure*}

We start our discussion on excited states with the electronic spectra of the \ce{UO_2^{2+}} cation which will be later compared to its isoelectronic NUN and \ce{PaO2^2+} analogs summarized in Table~\ref{tbl:adiabatic-spectrum}. 
The adiabatic electronic spectrum of the \ce{ThO2} species will be discussed separately as this molecule bears a different point group symmetry ($C_{2v}$ vs. $D_{\infty h}$) and its electronic excitations are of different character.   
First, we analyze our reference adiabatic excited states obtained with the CR-EOM-CCSD(T) method.  
As shown in Table~\ref{tbl:adiabatic-spectrum}, the lowest part of the singlet excited-state spectrum of \ce{UO_2^{2+}} is composed of excitations from the $\sigma_u$ (U 5f and U 6p) HOMO and the bonding $\pi_u$ (U 5f and O 2p) orbital to the unoccupied U 5f orbitals. 
Specifically for \ce{UO2^{2+}}, we focused on the following singlet electronically excited states: $\sigma_u \rightarrow \delta_u$, $\sigma_u \rightarrow \phi_u$,  $\pi_u \rightarrow \delta_u$, and $\pi_u \rightarrow \phi_u$. 
While the low-lying part of the electronic spectrum of NUN is very similar to \ce{UO_2^{2+}} in terms of their character, excitation energies are, however, lowered by 0.5 to 2 eV.  
As opposed to \ce{UO2^{2+}} and NUN, the lower part of the electronic spectrum of \ce{PaO2+} does not feature electronic transitions from the occupied $\pi_u$ orbital and the whole spectrum is shifted towards much higher excitation energies.
The $1^1\Phi_g$, $1^1\Delta_g$, and $1^1\Pi_g$ excited states characteristic for \ce{UO2^{2+}} are also present in NUN and \ce{PaO2+}.
The corresponding adiabatic excitation energies are lowered by about 1 eV for NUN and lifted up by about 1 eV for \ce{PaO2+} compared to the uranyl cation.
While the optimal bond lengths and vibrational frequencies for these excited states are similar in the \ce{UO_2^{2+}} and NUN molecules, the excited-state bond lengths in \ce{PaO2^+} are elongated by up to 0.1~\AA~and the corresponding vibrational frequencies lowered by 100-150 cm$^{-1}$.
Furthermore, the \ce{UO_2^{2+}} and NUN species feature one additional excited state, namely $2^1\Phi_g$, that is missing in the lower part of the energy spectrum of \ce{PaO2^+}. 
A common characteristic of the electronic spectra of NUN and \ce{PaO2^+} is the presence of the $1^1\Sigma_u$ and $1^1\Delta_u$ excited states in both molecules. 
Despite their identical orbital character, $\sigma_u \rightarrow \sigma_g$ and $\sigma_u \rightarrow \delta_g$ for $1^1\Sigma_u$ and $1^1\Delta_u$, respectively, the corresponding excitation energies differ by 2 and 1 eV, respectively.
Moreover, the differences in vibrational frequencies for these two states amount to 200 cm$^{-1}$. 
Finally, the electronic spectrum of the \ce{ThO2} molecule does not resemble the spectra of the other isoelectronic analogs. 
Specifically for \ce{ThO2} the investigated singlet excitation energies occur from the occupied O 2p and Th 6p orbitals to the unoccupied Th 6s orbital in the lower part of the spectrum and from the mixed O 2p and Th 6p/5d orbitals to the Th 7p/6d orbitals in the higher part of the spectrum.  

\begin{table*}[h!]
	\centering
	\small
	\caption{Vertical excitation energies of the lowest-lying singlet excited states of CUO and \ce{NUO+} for all investigated EOM-CC methods. All excitation energies are calculated for fixed molecular geometries: $r_{\rm U-C} = 1.733$ \AA{} and $r_{\rm U-O} = 1.779$ \AA{} for CUO and $r_{\rm U-N} = 1.698$ \AA{} and $r_{\rm U-O} = 1.761$ \AA{} for \ce{NUO+}, respectively. The differences with respect to CCSD/CR-EOM-CCSD(T) are given in parentheses.}\label{tbl:cuo-nuo+}
\begin{tabular}{llccccccc}
  & Method  & \multicolumn{7}{c}{\Tv\ [eV]} \\ \hline
  && {$1^1\Delta(\sigma\rightarrow\delta)$} 
  & {$1^1\Sigma(\sigma\rightarrow\sigma)$}
  & {$2^1\Delta(\sigma\rightarrow\delta)$}
  & {$1^1\Gamma(\pi\rightarrow\phi)$}
  & {$1^1\Phi(\sigma\rightarrow\phi)$}
  & {$2^1\Phi(\pi\rightarrow\delta)$}
  & {$1^1\Pi(\pi\rightarrow\sigma)$} \\\cline{3-9}
\multirow{5}{*}{CUO}
&EOM-pCCD+S    &  2.16($+$0.50)   & 2.82($+$1.22)   & 3.11($+$0.68)  & 4.01($+$1.64)   & 1.78($+$0.29)  & 3.31($+$0.88)  & 2.75($+$0.47) \\
&EOM-pCCD-CCS  &  2.17($+$0.51)   & 2.81($+$1.21)   & 3.10($+$0.67)  & 4.02($+$1.65)   & 1.79($+$0.30)  & 3.31($+$0.88)  & 2.75($+$0.47) \\
&EOM-pCCD-LCCSD&  1.59($-$0.07)   & 1.11($-$0.49)   & 2.13($-$0.30)  & 2.91($+$0.54)   & 1.43($-$0.06)  & 2.86($+$0.43)  & 2.61($+$0.33) \\
&EOM-CCSD      &  1.55($-$0.11)   & 1.97($+$0.37)   & 2.53($+$0.10)  & 2.08($-$0.29)   & 1.34($-$0.15)  & 2.18($-$0.25)  & 2.47($+$0.19) \\
&CR-EOM-CCSD(T)&  1.66 &            1.60            & 2.43           & 2.37            & 1.49           & 2.43           & 2.28          \\
\\
& & {$1^1\Delta(\sigma\rightarrow\delta)$}
& {$1^1\Gamma(\pi\rightarrow\phi)$}
& {$2^1\Delta(\pi\rightarrow\phi)$}
& {$1^1\Phi(\sigma\rightarrow\phi)$}
& {$1^1\Pi(\pi\rightarrow\delta)$} \\\cline{3-7}
\multirow{5}{*}{\ce{NUO+}}
&EOM-pCCD+S      & 3.65($+$0.84)   & 5.95($+$2.44)  & 6.22($+$2.51)   & 3.15($+$0.47)  & 5.47($+$1.65)  \\
&EOM-pCCD-CCS    & 3.68($+$0.87)   & 5.99($+$2.48)  & 6.26($+$2.55)   & 3.17($+$0.49)  & 5.51($+$1.69)  \\
&EOM-pCCD-LCCSD  & 3.04($+$0.23)   & 4.06($+$0.52)  & 4.22($+$0.52)   & 2.73($+$0.05)  & 4.03($+$0.21)  \\
&EOM-CCSD        & 2.81            & 3.54           & 3.71            & 2.52($-$0.15)  & 3.51($-$0.31)  \\ 
&CR-EOM-CCSD(T)  & --              &--              &--               & 2.68           & 3.82           \\
\end{tabular}
\end{table*}

The spectroscopic parameters obtained from the investigated EOM-pCCD-based methods and from the EOM-CCSD approach are collected in Table~\ref{tbl:adiabatic-spectrum}, while their corresponding deviations from the CR-EOM-CCSD(T) reference parameters are given in parentheses. 
In general, the simplest (and simultaneously the cheapest) EOM-pCCD+S and EOM-pCCD-CCS models overestimate adiabatic excitation energies by up to 3.5 eV for \ce{UO_2^{2+}}, up to 2.5 eV for NUN, up to 1.2 eV for \ce{PaO2+}, and up to 2.3 eV for \ce{ThO2}. 
These large errors compared to CR-EOM-CCSD(T) are significantly reduced by the EOM-pCCD-LCCSD approach, which gives comparable results to EOM-CCSD.
We should note that for all the targeted excited states summarized in Table~\ref{tbl:adiabatic-spectrum}, contributions form triple excitations seem to be negligible.
Similar observations can be made for \re\ and \we{}.
Specifically, the simple EOM-pCCD+S and EOM-pCCD-CCS methods considerably underestimate equilibrium bond lengths (up to 0.06 \AA{}) and generally overestimate vibrational frequencies (up to 120 cm$^{-1}$).
The errors in spectroscopic constants significantly reduce if dynamical correlation is included on top of pCCD.
In general, EOM-pCCD-LCCSD overestimates \re\ (differences typically amount to 0.03 \AA{} or less) and underestimates \we{}.
Although the conventional EOM-CCSD approach provides equilibrium bond lengths and vibrational frequencies that deviate less from CR-EOM-CCSD(T) reference data, EOM-pCCD-LCCSD predicts excitation energies closest to CR-EOM-CCSD(T) reference results.
A statistical analysis of adiabatic excitation energies \Ta\ (mean error (ME) and standard deviation) for all investigated molecules confirms that both EOM-pCCD-LCCSD and EOM-CCSD are of acceptable accuracy, with a mean error of approximately 0.05 eV and a standard deviation of at most 0.5 eV (see Figure~\ref{fig:errors}(a); the individual error measures are collected in Tables S14 and S15 of the ESI\dag).
Most importantly, EOM-pCCD-LCCSD on average outperforms EOM-CCSD in predicting (adiabatic) excitation energies, especially for the doubly-excited states present in the ThO and ThS molecules, reducing the standard deviation to 0.25 eV.
The mean errors for all excitation energies along the potential energy surface (calculated from eq.~\eqref{eq:me}) are displayed in Figure~\ref{fig:me_surface}.
In general, EOM-pCCD+S and EOM-pCCD-CCS result in unacceptably large ME, typically much larger than 1 eV, while EOM-CCSD and EOM-pCCD-LCCSD feature a ME of 0.5 eV or smaller.

\section{Vertical excitation energies and their statistical analysis}\label{sec:nuo-cuo}
In this section, we extend the isoelectronic series discussed above by two uranyl isoelectronic analogs, the CUO~\cite{cuo_93,cuo_acie,cuo_cej,cuo_nb_science,cuo_nb_jacs,cuo_ivan,cuo_ci,cuo_laura,cuo_nb_ic,cuo_2012,cuo_dmrg} and \ce{NUO^{+}}~\cite{xuy,pawel1,tecmer2014} molecules. 
Their vertical electronic transitions are collected in Table~\ref{tbl:cuo-nuo+}. 
We have chosen optimized equilibrium bond lengths of ${r_{\rm U-C}}$=1.733~\AA~and ${r_{\rm U-O}}$=1.779~\AA~for CUO~\cite{pawel3,cuo_dmrg} and ${r_{\rm U-N}}$=1.698~\AA~and ${r_{\rm U-O}}$=1.761~\AA~for \ce{NUO^{+}},~\cite{pawel1,tecmer2014} respectively.  
In the CUO molecule, the lowest-lying excitation energies occur from the $\sigma$ HOMO composed mainly from U 5f and from the $\pi$ molecular orbital with mixed U 6d/5f contributions to the unoccupied $\sigma$, $\delta$, and $\phi$ molecular orbitals. 
Although the character of some excited states is similar to those observed for \ce{UO2^{2+}} and NUN, a distinct feature of CUO is the low-lying transition to the $\sigma$ molecular orbital originating from the U 7s orbital. 
Similar to CUO, the electronic excitations in \ce{NUO^+} take place from the $\sigma$ HOMO and form the $\pi$ molecular orbitals with leading contributions from the U 5f and N 2p orbitals. 
In the lowest-lying vertical transitions, electrons are transferred to the unoccupied atomic U 5f orbitals ($\delta$ and $\phi$).  

The electronic spectra of CUO is rather dense, with excited states lying close in energy to each other (see Table~\ref{tbl:cuo-nuo+}).
Specifically, CR-EOM-CCSD(T) predicts excitations energies between 1.5 eV and 2.4 eV.
As expected, both EOM-pCCD+S and EOM-pCCD-CCS significantly overestimate excitation energies (errors lie between 0.3 eV to 1.7 eV), while EOM-pCCD-LCCSD yields excitation energies similar to EOM-CCSD.
Large deviations from CR-EOM-CCSD(T) reference data can be found for the $\sigma\rightarrow\sigma$ excitation, which is strongly underestimated by EOM-pCCD-LCCSD (almost 0.5 eV).
Furthermore, EOM-pCCD-LCCSD predicts this $1^1\Sigma(\sigma\rightarrow\sigma)$ state to be the first excited state, while it is the second lowest-lying excited state in CR-EOM-CCSD(T).
Overall, both EOM-CCSD and EOM-pCCD-LCCSD provide a different order of states in the lower part of the electronic spectrum compared to CR-EOM-CCSD(T) reference data.
We should highlight that both the CUO and \ce{NUO^{+}} molecules are rather complex test cases as their corresponding electronic spectra are difficult to model computationally.
Nonetheless, the errors in excitation energies predicted by EOM-pCCD-LCCSD are at most 0.5 eV, while the largest error encountered in EOM-CCSD calculations amounts to approximately 0.4 eV.

The lower part of Table~\ref{tbl:cuo-nuo+} summarizes the excitation energies for the \ce{NUO^{+}} molecule.
We should note that we encountered convergence difficulties in CR-EOM-CCSD(T) calculations and only two roots could be optimized.
Thus, EOM-CCSD results are taken as reference values in cases CR-EOM-CCSD(T) data is missing.
Most importantly, we observe similar trends in excitation energies as found for the remaining uranyl isoelectronic analogs.
EOM-pCCD+S and EOM-pCCD-CCS considerably overestimate excitation energies, while EOM-pCCD-LCCSD and EOM-CCSD predict excitation energies that agree well with CR-EOM-CCSD(T) reference data (with differences of 0.2 eV for EOM-pCCD-LCCSD and 0.3 eV for EOM-CCSD).
Furthermore, EOM-pCCD-LCCSD overrates excitation energies, while EMO-CCSD underestimates them with respect to CR-EOM-CCSD(T) data.
We should note that the differences in excitation energies between EOM-pCCD-LCCSD and EOM-CCSD are consistently about 0.2 and 0.5 eV, respectively.
Thus, we can anticipate that the corresponding errors with respect to (the missing) CR-EOM-CCSD(T) reference data will be of similar order, namely 0.2-0.3 eV.
Finally, a statistical analysis of vertical excitation energies \Tv\ (mean error (ME) and standard deviation) for all investigated molecules is presented in Figure~\ref{fig:errors}(b).
For ThO, ThS, \ce{UO2^{2+}}, NUN, \ce{PaO2^{+}}, and \ce{ThO2}, the vertical excitation energies have been determined with respect to the ground state equilibrium geometries obtained from various CC models (see also Tables~\ref{tbl:thoths} and \ref{tbl:adiabatic-spectrum}). 
The error measures confirm that both EOM-pCCD-LCCSD and EOM-CCSD are of acceptable accuracy, with a mean error of approximately 0.05 eV and a standard deviation of 0.3 eV (EOM-pCCD-LCCSD) and 0.5 eV (EOM-CCSD), respectively.

\section{Conclusions and outlook}\label{sec:conclusions}
In this work, we have investigated the ground and singlet excited states of model di- and tri-atomic f0 actinide species including the ThO, ThS, \ce{UO^{2+}_2}, \ce{NUO^{+}}, CUO, \ce{NUN}, \ce{PaO^{+}_2}, and \ce{ThO_2} molecules employing different variants of the EOM-CC formalism.
Specifically, we investigated various pCCD-based models (EOM-pCCD+S, EOM-pCCD-CCS, and EOM-pCCD-LCCSD) as well as the conventional EOM-CCSD and the CR-EOM-CCSD(T) variants.
Our study shows that the EOM-pCCD-LCCSD method provides a reliable and accurate alternative to standard EOM-CC flavors. 
Specifically, compared to CR-EOM-CCSD(T) reference data, EOM-pCCD-LCCSD gives the smallest mean error (about 0.05 eV) and standard deviation (about 0.25 eV) for excitation energies in our actinide test set, reducing the standard deviation of EOM-CCSD (0.5 eV) by approximately a factor of 2.

Furthermore, our analysis highlights the marginal importance of triple excitations within the CR-EOM-CCSD(T) formalism.
The only exceptions are the doubly excited states in the ThO and ThS species. 
For these particular states, EOM-pCCD-LCCSD is clearly superior to EOM-CCSD, reducing the corresponding errors in excitation energies from 1.6 eV (EOM-CCSD) to 0.4 eV (EOM-pCCD-LCCSD). 
While the EOM-pCCD-LCCSD method provides more accurate excitation energies featuring the smallest error measures compared to CR-EOM-CCSD(T) reference data, spectroscopic constants (\re{} and \we{}) are better described by EOM-CCSD. 
In general, the EOM-pCCD-LCCSD optimal bond lengths are overestimated, while the calculated vibrational frequencies are usually underestimated.

{Although being a simplification of the conventional EOM-CCSD method, EOM-pCCD-LCCSD, on average, improves the description of excitation energies.
Yet, this decent and satisfactory performance of the EOM-pCCD-LCCSD approach might be attributed to error cancellations due to, for instance, the linear ansatz of the CCSD correction on top of pCCD.
A detailed analysis of the quality of the LCCSD correction and the resulting pCCD-LCCSD wavefunction as well as different tailored coupled cluster flavours are currently being investigated in our laboratory.
}
In order to further reduce errors in excitation energies and spectroscopic constants, the EOM-pCCD-LCCSD approach can be systematically improved.
One possibility is to employ the frozen pair CC (fpCCD~\cite{frozen-pccd}) model in the EOM formalism instead of its linearized version, LCCSD.
Furthermore, an orbital optimization protocol for excited states might provide a better description of spectroscopic constants of excited states. 
Moreover, the simplified EOM-pCCD-based models can be significantly improved by including triple excitations in some approximate manner in the pCCD correction for ground and excited states or introducing an active space triple correction for excited states.~\cite{kowalski2000active,eomcc_2001,kowalski2010active}
{Finally, open-shell electronic structures can be targeted using the conventional toolbox of coupled cluster theory.
For instance, triplet excited states can be optimized by considering an (explicit spin coupled) triplet excitation space, using the framework of linear response coupled cluster theory\mbox{\cite{lr-cc-triplet-1,lr-cc-triplet-2}} or the equation of motion formalism.\mbox{\cite{bartlett_2007}}}
All these flavours are currently being developed in our laboratory.  

\section{Acknowledgement}\label{sec:acknowledgement}
A.N.~acknowledges financial support from a SONATA BIS grant of the National Science Centre, Poland (no.~2015/18/E/ST4/00584).
P.T.~thanks a POLONEZ 1 research grant financed by Marie-Sk\l{}odowska--Curie COFUND. This project has received funding from the European Union's Horizon 2020 research and innovation programme under the Marie Skłodowska--Curie grant agreement No 665778. 
K.B.~acknowledges a Marie-Sk\l{}odowska-Curie Individual Fellowship project no.~702635--PCCDX, and a scholarship for outstanding young scientists from the Ministry of Science and Higher Education.

Calculations have been carried out using resources provided by Wroclaw Centre for Networking and Supercomputing (http://wcss.pl), grant nos.~411 and~412.


\normalem
\providecommand*{\mcitethebibliography}{\thebibliography}
\csname @ifundefined\endcsname{endmcitethebibliography}
{\let\endmcitethebibliography\endthebibliography}{}

\end{document}